\documentclass[twocolumn,prl]{revtex4}
\usepackage{epsf}
\usepackage{amsmath,amsthm,amssymb}
\usepackage{color}
\usepackage{dcolumn}
\usepackage{bm}
\usepackage{graphicx,epsfig}

\graphicspath{{fig/}}
\everymath{\displaystyle}

\begin{document}           

\title{Manipulating twisted electron beams}

\author{Alexander J. Silenko} \email{alsilenko@mail.ru}
\affiliation{Institute of Modern Physics, Chinese Academy of Sciences, Lanzhou 730000, China}
\affiliation{Bogoliubov Laboratory of Theoretical Physics, Joint Institute for Nuclear Research, Dubna 141980, Russia}
\affiliation{Research Institute for Nuclear Problems, Belarusian State University, Minsk 220030, Belarus}

\author{Pengming Zhang}
\email{zhpm@impcas.ac.cn} \affiliation{Institute of Modern Physics, Chinese Academy of Sciences, Lanzhou 730000, China}
\affiliation{University of Chinese Academy of Sciences, Yuquanlu 19A, Beijing 100049, China}

\author{Liping Zou}
\email{zoulp@impcas.ac.cn} \affiliation{Institute of Modern Physics, Chinese Academy of Sciences, Lanzhou 730000, China}
\affiliation{University of Chinese Academy of Sciences, Yuquanlu 19A, Beijing 100049, China}

\date{\today}

\begin {abstract}
A theoretical description of twisted (vortex) electrons interacting with electric and magnetic fields is presented, based on Lorentz transformations. The general dynamical equations of motion of a twisted electron with an intrinsic orbital angular momentum in an external field are derived. Methods for the extraction of an electron vortex beam with a given orbital polarization and for the manipulation of such a beam are developed.
\end{abstract}


\maketitle


The discovery of twisted (vortex) electron beams carrying intrinsic orbital angular momentum (OAM) \cite{UTV} has proven the existence of vortex states of a free electron. Unlike corresponding light vortex beams (which have been successfully used for 25 years \cite{Allen}), electron beams are charged. Therefore, they also possess significant orbital magnetic moments. Amazingly, a vortex electron in vacuum can be described by the usual Dirac equation for a free particle.

In the present work, the system of units $\hbar=1,~c=1$ is used. We include $\hbar$ and
$c$ explicitly when this inclusion clarifies the problem. We use the weak-field approximation and neglect terms quadratic in external fields.

We give a detailed classical description of dynamics of a twisted particle in external electromagnetic fields. There exists the perfect agreement between \emph{relativistic} equations of motion for the momentum and the spin in classical electrodynamics and quantum mechanics of spin-1/2 particles in electromagnetic fields (see Refs. \cite{RPJ,RPJSTAB,GBMT,JINRLet1,PRAnonstat,PhysScr,EPJC2017} and references therein). The wonderful agreement with the corresponding classical 
equations takes place for relativistic spin-1/2 particles in gravity \cite{gravity}. Relativistic equations of motion for spin-0 \cite{otherspins} and spin-1 \cite{PRDspin} particles also fully agree with the corresponding classical equations. This means that the use of an appropriate classical approach for obtaining equations of motion is perfectly admissible.

From the viewpoint of quantum mechanics, a twisted electron is a single pointlike particle. Its wave function mirrors a density of a probability to find the electron in a given point of the space. The standard classical model of electron in an atom developed by founders of quantum mechanics is an electron cloud \cite{wiki} characterizing a spatial distribution of an electron charge. When the atomic electron has a nonzero OAM, the model of the
rotating charged cloud is used. We adopt this model to the considered problem. A rotation of the charged electron cloud is a classical counterpart of a current operator describing a motion of the electron about the direction of the intrinsic OAM.
%
In this simple classical picture, an intrinsic OAM originating from the cloud rotation can be parallel to the momentum direction and can be nonzero for a particle at rest. So, the classical description should use some intrinsic rotation which is not associated with the electron momentum $\bm p$. Besides the intrinsic rotation, an extrinsic rotation of the electron can take place (for example, in an external magnetic field). The latter rotation depends on the electron momentum and is defined by the extrinsic OAM $\bm r\times\bm p$. Quantum mechanics uses the single operator of the OAM defined by $-i\hbar\bm r\times\nabla$. Despite the orthogonality of the classical quantities $\bm p$ and $\bm r\times\bm p$, the \emph{expectation values} of the operators $-i\hbar\nabla$ and $-i\hbar\bm r\times\nabla$ can be parallel to each other. The intrinsic and extrinsic OAMs have been introduced in Ref. \cite{BliokhSOI}.

 The simple classical model of the
rotating charged cloud was not previously used because precedent investigations followed the quantum-mechanical approach.
There are several quantum-mechanical descriptions of electron vortex states as axially symmetric plane waves (see reviews \cite{BliokhSOI,Lloyd}). A standard approach consists in the use of the paraxial approximation \cite{Siegman}. In this approximation, $|p_x|\ll p,\,|p_y|\ll p$ when a wave moves close to the $z$ axis. Another solution of the wave equation is expressed in terms of the Bessel functions  \cite{BliokhSOI,Bliokhmagnetic}:
\begin{equation}
\psi_l^B\sim J_{|l|}(k_\bot\rho)\exp{[i(l\phi+k_zz)]},\qquad k_\bot^2+k_z^2=k^2.
\label{Bssel}\end{equation}
Here $J_l$ is the Bessel function of the first kind, $l=0,\pm1,\pm2,\dots$ is the azimuthal quantum
number, $k_z = p_z/\hbar$ is the longitudinal wave number, $k_\bot = p_\bot/\hbar$ is the transverse (radial) wave number, and $k=p/\hbar$. One more description results in the use of exponential wave packets of Dirac electrons \cite{Bialynicki-Birula}.

The wave function (\ref{Bssel}) is the most relevant to the goals of the present study because the paraxial approximation may become invalid after a Lorentz transformation. Indeed, the free electron can be considered to be in its rest frame. The Lorentz transformation of the OAM from the lab frame ($\bm L=L_z\bm e_z$) to the rest frame results in $\bm L^{(0)}=\bm L$.
The OAM in the frame moving with the arbitrary velocity $\bm V$ relative to the particle rest frame is given by
\begin{equation}\begin{array}{c}
\bm L=\frac{\epsilon}{mc^2}\bm L^{(0)}-\frac{(\bm L^{(0)}\cdot\bm p)\bm p}{m(\epsilon+mc^2)}, \qquad \epsilon=\frac{mc^2}{\sqrt{1-\frac{{\bm V}^2}{c^2}}}.
\end{array} \label{oamoper} \end{equation}
The orbital helicity of the particle is equal to
\begin{equation}\begin{array}{c}
h_{orb}\equiv\bm L\cdot\bm e=\bm L^{(0)}\cdot\bm e,
\end{array} \label{orbhl} \end{equation} where $\bm e=\bm p/p$ is the unit vector parallel to the momentum direction.

Equations (\ref{oamoper}) and (\ref{orbhl}) prove that the mutual orientation of the momentum and the OAM and the orbital helicity of the vortex Dirac particle depend on a reference frame. So, equation (\ref{Bssel}) for the Bessel wave is preferable. In this case, the $z$ axis can be attributed not only to the momentum direction but also to any other direction.

To develop methods for the manipulation of electron vortex beams, we need to determine the evolution of the OAM in electromagnetic fields. This determination will be based on Lorentz transformations. These transformations process several critical properties for the intrinsic and extrinsic orbital angular momenta as compared with the spin.

While the commutation relations and the Poisson brackets for the OAM and the spin are very similar, there is one big difference between the two quantities. The OAM is formed by the spatial components of the antisymmetric tensor $L^{\mu\nu}\equiv x^{\mu}p^{\nu}-x^{\nu}p^{\mu}$. Unlike the OAM, the conventional spin $\bm\zeta$ is defined by the spatial part of the four-component spin pseudovector $a^\mu$ in the particle rest frame. The connection between the four-component spin pseudovector and the antisymmetric spin tensor is given by (see, e.g., Ref. \cite{Landau4}) $a^\lambda=-2e^{\lambda\mu\nu\rho}\mathfrak{S}_{\mu\nu}u_\rho$, where $e^{\lambda\mu\nu\rho}$ is the
four-dimensional completely antisymmetric unit tensor (Levi-Civita tensor), $u^\rho=p^\rho/m$ is the four-velocity and $a^\mu u_\mu=0$. The spatial components of the spin tensor $\mathfrak{S}^{\mu\nu}$ form the three-component spatial pseudovector $\bm{\mathfrak{S}}$ which is not equivalent to $\bm\zeta$.

The key reason for the decomposition of the OAM into intrinsic and extrinsic parts is its nonzero value in the particle rest frame ($\bm p=0$). This value defines the intrinsic OAM originating from the rotation of the charged cloud. The extrinsic OAM is defined by the motion of the center of charge of the electron. Evidently, we can suppose this motion to be independent of the presence of the intrinsic OAM. As a result, the dynamics of the two parts of the electric dipole moment (EDM) can be independently described.

We need to repeat that the contributions of the intrinsic and extrinsic parts of the OAM into the quantum-mechanical Hamiltonian are described by a single operator, $-i\hbar\bm r\times\nabla$.

Another key point is the difference between dynamics of the intrinsic OAM and the spin. This difference is caused by different definitions of the two quantities. Since the conventional three-component spin is defined in the particle rest frame being accelerated in external fields, the angular velocity of its precession includes the correction for the Thomas effect. Contrary to this, the OAM is defined via the antisymmetric tensor $L^{\mu\nu}$ (see above) and the angular velocity of its precession does not include this correction. For a relativistic description, it is convenient to consider the nonrotating instantaneous inertial frame accompanying the particle. In this frame, $\bm p=0$ at the given moment of time. The magnetic moment caused by the orbital motion of the charged cloud is equal to
\begin{equation}\begin{array}{c}
\bm\mu^{(0)}=\frac{e}{2mc}\bm L^{(0)}.
\end{array} \label{meqfnia} \end{equation}
The corresponding relativistic quantum-mechanical formula has been derived in Ref. \cite{Barut}.
The angular velocity of the precession of the intrinsic OAM in the nonrotating instantaneous inertial frame is defined by the Larmor formula:
\begin{equation}\begin{array}{c}
\bm\Omega^{(0)}=-\frac{e}{2mc}\bm B^{(0)}.
\end{array} \label{Lmfl} \end{equation}

An interaction of the electric and magnetic dipole moments, $\bm d$ and $\bm\mu$, with the external fields is defined by the
general Hamiltonian
\begin{equation}   \begin{array}{c}
H=-\bm d\cdot\bm E-\bm\mu\cdot\bm B,
\end{array} \label{multexp} \end{equation}
where all quantities are defined in the lab frame. Now we can take into account that the quantities $\bm L^{(0)}$ and $\bm\mu^{(0)}$ are connected with the nonrotating instantaneous inertial frame and can perform the relativistic transformation of the dipole moments to the lab frame. It has the form \cite{PhysScr}
\begin{eqnarray} 
&& \bm d=\bm \beta\times\bm \mu^{(0)},\qquad
\bm\mu=\bm\mu^{(0)}-\frac{\gamma}{\gamma+1}\bm\beta(\bm\beta\cdot\bm\mu^{(0)}),
\qquad \nonumber \\
&& \bm\beta=\frac{\bm v}{c},\qquad  \gamma=(1-\bm\beta^2)^{-1/2}
\label{meqfinal}
\end{eqnarray}
where we have used the fact $\bm d^{(0)}=0$.

As a result, the Hamiltonian is given by
\begin{eqnarray}
H& = & -\dfrac{e}{2mc}\Big [\bm B\cdot\bm L^{(0)}-\frac{\gamma}{\gamma+1}(\bm\beta\cdot\bm B)(\bm\beta\cdot\bm L^{(0)})
\nonumber \\
&&-(\bm\beta\times\bm E)\cdot\bm L^{(0)}\Big ]
\label{emgaoHn}
\end{eqnarray}
or, using Eq. (\ref{oamoper}),
\begin{equation}\begin{array} {c}
H=-\frac{e}{2mc\gamma}\left[\bm B\cdot\bm L-(\bm\beta\times\bm E)\cdot\bm L\right].
\end{array}\label{emgaoHl}\end{equation}

The use of Poisson brackets allows us to derive the relativistic equation for the Larmor precession of the intrinsic OAM:
\begin{equation}\begin{array}{c}
\frac{d\bm L}{dt}=\bm\Omega\times\bm L,\qquad \bm\Omega=-\frac{e}{2mc\gamma}\left[\bm B-\bm\beta\times\bm E\right].
\end{array} \label{reLarpr} \end{equation}

It should be added that the torque provided by the electric field is nonzero only in the presence of the EDM $\bm d$, but it originates only from a motion of the magnetic dipole moment.

We expect that the relativistic equation for the Larmor precession of the extrinsic OAM is similar. When spin effects are disregarded, the Hamiltonian of a particle in a uniform magnetic field takes the form
\begin{equation}\begin{array}{c}
H=\sqrt{m^2+(\bm p-e\bm A)^2}, \qquad \bm A=\frac12\bm B\times\bm r.
\end{array} \label{Hamilfn} \end{equation}
In the weak-field approximation,
\begin{equation}\begin{array}{c}
H=\sqrt{m^2+\bm p^2}-\frac{e\bm B\cdot\bm L^{(e)}}{2\sqrt{m^2+\bm p^2}}
\end{array} \label{Hamiltf} \end{equation}
and the Larmor precession of the extrinsic OAM in the uniform magnetic field is given by
\begin{eqnarray}
 \dfrac{d\bm L^{(e)}}{dt}&=&\bm\Omega^{(e)}\times\bm L^{(e)},\qquad \nonumber \\
 \bm\Omega^{(e)}&=&-\dfrac{e\bm B}{2\sqrt{m^2+\bm p^2}}=-\dfrac{e}{2m\gamma}\bm B.
\label{reLarpe}
\end{eqnarray}
This equation is similar to Eq. (\ref{reLarpr}).

It is instructive to compare the results for the intrinsic OAM and the spin. The use of the well-known Lorentz transformations for the fields brings the \emph{lab frame} Hamiltonian to the form
\begin{equation}\begin{array} {c}
H=-\frac{e}{2mc}\bm B^{(0)}\cdot\bm L^{(0)}.
\end{array}\label{emgapHn}\end{equation}
The \emph{lab frame} Hamiltonian for the spinning particle in the absence of the EDM is given by
($|\bm\zeta|=s$) \cite{PhysScr}
\begin{equation}   \begin{array}{c}
H=-\frac{e}{mc}\left(\frac{g\bm B^{(0)}}{2\gamma}+\frac{\bm\beta\times\bm E^{(0)}}{\gamma+1}\right)\cdot\bm\zeta,
\end{array} \label{multred} \end{equation} where $g=2mc\mu/(es)$ and $s$ is the spin quantum number.
Evidently, the main difference between Eqs. (\ref{emgapHn}) and (\ref{multred}) is the absence of Thomas precession for the intrinsic OAM.

Our result for the dynamics of the intrinsic OAM in an electric field differs from that presented (without derivation) in Refs. \cite{BliokhSOI,Bliokh2007}. The equation (2.27) in Ref. \cite{BliokhSOI} and the equation (8) in Ref. \cite{Bliokh2007} cannot be correct because they predict an infinitely large angular velocity of precession of the intrinsic OAM at $p\rightarrow0$. Only the particle momentum direction shows such a behavior, see Ref. \cite{RPJSTAB}.

The results obtained permit us to develop methods for the manipulation of electron vortex beams. These methods are similar to those governing the spin while formulas defining dynamics of the intrinsic OAM and the spin differ.
We can specify manipulations of the electron vortex beams and present their quantitative descriptions.

\emph{Separation of beams with opposite directions of the OAM.}--- Such a separation can be achieved in a longitudinal magnetic field. This field can be nonuniform \cite{BliokhSOI} and even uniform. The nonuniform longitudinal magnetic field leads to a force acting on the OAM. Because of the one-to-one correspondence between the OAM and the corresponding magnetic moment, the direction of this force depends on that of the OAM. As a result, accelerations of particles with oppositely directed OAMs have different signs. This leads to different velocities of particles with the opposite directions of the OAMs. Therefore, the beam with a given OAM direction can be extracted (e.g., with the Wien filter).

Importantly, even a uniform longitudinal magnetic field leads to a dependence of a particle velocity on the OAM direction. If particles have equal energies beyond the longitudinal magnetic field, their velocities in this field satisfy the following equation:
\begin{equation}\begin{array} {c}
v=v_0+\frac{e}{2m^2c^2\beta_0\gamma_0^4}\bm B\cdot\bm L,
\end{array}\label{eveln}\end{equation}
where $\bm v_0$ is the particle velocity beyond the field. This effect either decreases or increases the beam separation caused by the \emph{nonuniform} longitudinal magnetic field.

The manipulations considered below allow one to extract a beam with a needed orbital polarization.

The use of a transversal magnetic field is less convenient because this field can bring a lost of beam coherence as a result of Larmor precession.

\emph{Freezing the intrinsic OAM in electromagnetic fields.}--- As in spin physics \cite{FJM}, it is important to consider a condition which allows one to freeze the intrinsic OAM (i.e., to keep the orbital helicity constant) in electromagnetic fields. These fields deflect the beam. We consider a potential for a beam deflection without a change of $h_{orb}$ defined by Eq. (\ref{orbhl}). In this case, the angular velocity of the relativistic Larmor precession (\ref{reLarpr}) should be equal to the angular velocity of the rotation of the momentum direction $\bm N=\bm p/p\approx\bm\beta/\beta$ \cite{RPJSTAB}:
\begin{equation}\frac{d\bm
N}{dt}=\bm\omega\times\bm N, \qquad \bm\omega=-\frac{e}{mc\gamma}\left(\bm B-\frac{\bm N\times\bm E}{\beta}\right).
\label{eqm}\end{equation} The standard geometry is $\bm E\bot\bm B\bot\bm\beta$.

The condition $\bm\Omega_L=\bm\omega$ is satisfied when
\begin{equation} \bm B=\left(\frac{2}{\beta^2}-1\right)\bm\beta\times\bm E.
\label{eqBA}\end{equation}
The device defined by Eq. (\ref{eqBA}) is a deflector of the electron vortex beams which freezes the OAM relative to the momentum direction. It rotates the beam direction with the angular velocity
\begin{equation} \bm\Omega_L=\bm\omega=-\frac{e\bm B}{mc\gamma(\gamma^2+1)}.
\label{danvl}\end{equation}
For standard beams \cite{VGRILLO} with energy of the order of $10^2$ keV, the deflection is rather effective.

We should add that the use of the proposed beam deflector after or together with the longitudinal magnetic field considered above allows one to separate out electrons with oppositely directed OAMs.

\emph{Rotator of the intrinsic OAM.}--- An other important beam manipulation is a rotation of the intrinsic OAM relative to the momentum direction. This manipulation is desirable and even necessary when the beam is confined in a storage ring or trap. In this case, it is convenient to direct the OAM of the beam vertically. The vertical direction is preferable because it is not affected by the main vertical magnetic field and this orbital polarization can be conserved.  In accelerator physics, one frequently uses a Wien filter as a spin rotator. This device can also be applied as a OAM rotator. In this case, $\bm E\bot\bm B\bot\bm\beta$ and the Lorentz force acting on electrons is equal to zero ($\bm E=-\bm\beta\times\bm B$). Therefore, Eq. (\ref{reLarpr}) for the relativistic Larmor precession takes the form
\begin{equation}\begin{array}{c}
 \bm\Omega^{(W)}=-\frac{e}{2mc\gamma^3}\bm B.
\end{array} \label{reLW} \end{equation}

When beams with oppositely directed OAMs have different velocities, a Wien filter allows one to extract electrons with one of orbital polarizations.

\emph{Flipping the intrinsic OAM.}--- If an electron vortex beam with a upward or a downward orbital polarization is confined in a storage ring, the direction of the intrinsic OAM can be flipped. A flip of the OAM is similar to that of the spin and can be fulfilled by the method of the magnetic resonance. A significant difference between the flips of the OAM and the spin consists in different dependences of the resonance frequencies on the electric and magnetic fields. A spin flip frequency in a storage ring is defined by
the Thomas-Bargmann-Mishel-Telegdi equation (see Ref. \cite{EPJC2017} for details) whose distinction from Eq. (\ref{reLarpr}) is evident. The OAM flip can be forced by a longitudinal (azimuthal) magnetic field oscillating with the resonance frequency. A Wien filter with a vertical electric field and a radial magnetic field oscillating with the resonance frequency \cite{EPJC2017,WienFil} can also be used for the OAM flip.

In summary, we have presented the basic theoretical description of dynamics of the vortex electrons in electromagnetic fields. Our derivations have been based on the classical approach and Lorentz transformations. We have shown that the orbital polarization of such electrons can be managed and have developed basic methods for the manipulation of electron vortex beams. We expect that the results presented can be applied not only to electrons but also to other particles. In particular, twisted positron beams could also find important applications. We suppose that they could be used for testing magnetic materials and for a formation of twisted positronium atoms. For these purposes, a manipulation of twisted positron beams (for example, their deceleration) is necessary.

We believe that the obtained theoretical results can also be useful for high-energy physics.

In precedent theoretical investigations (see Refs.
\cite{BliokhSOI,Lloyd,Bliokhmagnetic,Bialynicki-Birula,Bliokh2007,Chowdhury,Hayrapetyan,RajabiBerakdar,Barnett} and references therein), the quantum-mechanical approach was used. However, it is very difficult to fulfill an
appropriate quantum-mechanical analysis for a twisted particle in \emph{general} electromagnetic
fields. As a result, previous publications focused an attention on magnetic or other specific interactions.
Otherwise, a consideration of only magnetic interactions allows one neither obtaining a general
relativistic equation of motion for the intrinsic OAM nor developing methods of a manipulation
of twisted electron beams. We should add that such a manipulation is impossible if only the
magnetic field is used. We present the solution of
the two problems. In our work, the general relativistic equation of motion for
the intrinsic OAM has been obtained and the methods of a manipulation of twisted electron beams
have been developed for the first time.

\emph{Acknowledgements.}---
This work was supported by the Belarusian Republican Foundation for Fundamental Research
(Grant No. $\Phi$16D-004), by the National Natural Science Foundation of China (Grant No. 11575254), and
by the Heisenberg-Landau program of the German Ministry for
Science and Technology (Bundesministerium f\"{u}r Bildung und
Forschung).
A.S. also acknowledges hospitality and support by the Institute of Modern Physics of the Chinese Academy of Sciences. The authors are grateful to D. Chowdhury and A. G. Hayrapetyan for paying their attention to Refs. \cite{Chowdhury,Hayrapetyan}  and to Jarah Evslin for helpful exchanges.


\begin{thebibliography}{}

\bibitem{UTV}
M. Uchida, A. Tonomura, Generation of electron beams carrying orbital angular momentum, Nature \textbf{464}, 737 (2010); 
J. Verbeeck, H. Tian, P. Schattschneider, Production and application of electron vortex beams, Nature \textbf{467}, 301 (2010). 

\bibitem{Allen}
L. Allen, M. W. Beijersbergen, R. J. C. Spreeuw, J. P. Woerdman, Orbital angular momentum of light and the transformation
of Laguerre-Gaussian laser modes, Phys. Rev. A \textbf{45}, 8185 (1992). 


\bibitem{RPJ}
A. J. Silenko, Quantum-mechanical description of
the electromagnetic interaction of relativistic particles
with electric and magnetic dipole moments, Russ. Phys. J. {\bf 48}, 748
(2005).


\bibitem{RPJSTAB}
A. J. Silenko, Equation of spin motion in storage rings in the cylindrical coordinate system,
Phys. Rev. ST Accel. Beams \textbf{9}, 
034003 (2006).


\bibitem{GBMT}
T. Fukuyama and A. J. Silenko, Derivation of Generalized
Thomas-Bargmann-Michel-Telegdi Equation for a Particle with
Electric Dipole Moment, Int. J. Mod. Phys. A \textbf{28}, 1350147
(2013).

\bibitem{JINRLet1}
A. J. Silenko, Classical Limit of Equations of the Relativistic Quantum Mechanics in the Foldy-Wouthuysen Representation, Phys. Part. Nucl. Lett. \textbf{10}, 91 
(2013).

\bibitem{PRAnonstat}
A. J. Silenko, Energy expectation values of a particle in nonstationary fields,
Phys. Rev. A \textbf{91}, 012111 (2015).


\bibitem{PhysScr}
A. J. Silenko, Spin precession of a particle with an electric
dipole moment: contributions from classical electrodynamics and
from the Thomas effect, Phys. Scripta \textbf{90}, 065303 (2015).

\bibitem{EPJC2017}
A. J. Silenko, General classical and quantum-mechanical description of magnetic resonance: an application to
electric-dipole-moment experiments, Eur. Phys. J. C \textbf{77}, 341 (2017).


\bibitem{gravity}
A. J. Silenko and O. V. Teryaev, Semiclassical limit for Dirac
particles interacting with a gravitational field,
Phys. Rev. D {\bf 71}, 064016 (2005);
A. J. Silenko, Classical and quantum spins in curved spacetimes,
Acta Phys. Polon. B Proc. Suppl. {\bf 1}, 87 (2008);
Yu. N. Obukhov, A. J. Silenko, and O. V. Teryaev,
Spin dynamics in gravitational fields of rotating bodies
and the equivalence principle,
Phys. Rev. D {\bf 80}, 064044 (2009); Dirac fermions in strong gravitational fields,
Phys. Rev. D {\bf 84}, 024025 (2011); Spin in an arbitrary gravitational field,
Phys. Rev. D {\bf 88}, 084014 (2013); Spin-torsion coupling and gravitational moments of Dirac
fermions: Theory and experimental bounds,
Phys. Rev. D {\bf 90}, 124068 (2014); Spin-Gravity Interactions and Equivalence Principle,
Int. J. Mod. Phys.: Conf. Ser. {\bf 40}, 1660081 (2016); Manifestations of the rotation and gravity of the Earth in high-energy physics experiments, Phys. Rev. D \textbf{94}, 044019 (2016); General treatment of quantum and classical spinning particles in
external fields, Phys. Rev. D (2017), accepted for publication.

\bibitem{otherspins}
A.\,J. Silenko, Hamilton Operator and the Semiclassical Limit
for Scalar Particles in an Electromagnetic Field,
Theor. Math. Phys. \textbf{156}, 
1308 (2008);
Scalar particle in general inertial and gravitational fields and conformal invariance revisited, Phys. Rev. D {\bf 88}, 045004 (2013);
New symmetry properties of pointlike scalar and Dirac particles, Phys. Rev. D {\bf 91}, 065012 (2015).

\bibitem{PRDspin} A. J. Silenko, Quantum-mechanical description of spin-1 particles with electric dipole moments,
Phys. Rev. D {\bf 87}, 073015 (2013).

\bibitem{wiki} Electron cloud, Wikipedia, https://simple.wikipedia.org/wiki/Electron$\_$cloud


\bibitem{BliokhSOI}
K. Y. Bliokh, I. P. Ivanov, G. Guzzinati, L. Clark, R. Van Boxem, A. B\'{e}ch\'{e}, R. Juchtmans,
M. A. Alonso, P. Schattschneider, F. Nori, and J. Verbeeck, Theory and applications of free-electron
vortex states, Phys. Rep. \textbf{690}, 1 (2017). 

\bibitem{Lloyd}
S. M. Lloyd, M. Babiker, G. Thirunavukkarasu, and J. Yuan, Electron vortices: Beams with orbital angular momentum, Rev. Mod. Phys. \textbf{89}, 035004 (2017).

\bibitem{Siegman}
 A. E. Siegman, \emph{Lasers}, (University Science Books, Mill Valley, 1986); M. B. Vinogradova, O. V. Rudenko, A. P. Sukhorukov, \emph{Wave theory}, (Nauka, Moscow, 1990) [in Russian].

\bibitem{Bliokhmagnetic}
K. Y. Bliokh, P. Schattschneider, J. Verbeeck, F. Nori, Electron vortex beams in a magnetic field: A new twist on Landau
levels and Aharonov-Bohm states, Phys. Rev. X \textbf{2}, 041011 (2012).

\bibitem{Bialynicki-Birula}
I. Bialynicki-Birula and Z. Bialynicka-Birula, Relativistic electron wave packets carrying
angular momentum, Phys. Rev. Lett. \textbf{118}, 114801 (2017).

\bibitem{Landau4}
V. B. Berestetskii, E. M. Lifshitz, L. P. Pitaevskii, \emph{Quantum Electrodynamics}, 2nd ed., (Pergamon Press, Oxford, 1982), p. 108.

\bibitem{Barut}
A. O. Barut and A. J. Bracken, Magnetic-moment operator of the relativistic electron, Phys. Rev.
D \textbf{24}, 3333 (1981).

\bibitem{Bliokh2007}
K. Y. Bliokh, Y. P. Bliokh, S. Savel’ev, F. Nori, Semiclassical dynamics of electron wave packet states with phase vortices,
Phys. Rev. Lett. \textbf{99}, 19040 (2007).

\bibitem{FJM}
F. J. M. Farley, K. Jungmann, J. P. Miller, W. M. Morse, Y. F.
Orlov, B. L. Roberts, Y. K. Semertzidis, A. Silenko, and E. J.
Stephenson, A new method of measuring electric dipole moments in
storage rings, Phys. Rev. Lett. \textbf{93}, 052001 (2004).

\bibitem{VGRILLO}
V. Grillo, G. C. Gazzadi, E. Karimi, E. Mafakheri, R. W. Boyd, and S. Frabboni, Highly Efficient Electron Vortex Beams Generated by Nanofabricated Phase Holograms, Appl. Phys. Lett. \textbf{104}, 043109 (2014); V. Grillo, G. C. Gazzadi, E. Mafakheri, S. Frabboni, E. Karimi, and R. W. Boyd, Holographic Generation of Highly Twisted Electron Beams, Phys. Rev. Lett. \textbf{114}, 034801 (2015).

\bibitem{WienFil}
W. M. Morse, Y. F. Orlov, and Y. K. Semertzidis, rf Wien filter in
an electric dipole moment storage ring: The partially frozen spin
effect, Phys. Rev. ST Accel. Beams \textbf{16}, 114001 (2013);
J. Slim, R. Gebel, D. Heberling, F. Hinder, D. H\"{o}lscher, A. Lehrach, B. Lorentz, S. Mey, A. Nass, F. Rathmann, L. Reifferscheidt, H. Soltner, H. Straatmann, F. Trinkel, J. Wolters,
Electromagnetic Simulation and Design of a Novel Waveguide RF Wien
Filter for Electric Dipole Moment Measurements of Protons
and Deuterons, Nucl. Instrum. Meth. A \textbf{828}, 116 (2016); 
A. Saleev \emph{et al.} (JEDI Collaboration), Spin tune mapping as a novel tool to probe the spin dynamics in storage rings, Phys. Rev. ST Accel. Beams \textbf{20}, 072801 (2017).


\bibitem{Chowdhury}
D. Chowdhury, B. Basu, P. Bandyopadhyay, Electron vortex beams in a magnetic field and spin filter, Phys. Rev. A \textbf{91}, 033812 (2015); P. Bandyopadhyay, B. Basu, D. Chowdhury, The geometric phase and the geometrodynamics of relativistic electron vortex beams, Proc. Roy. Soc. A \textbf{470}, 20130525 (2014); Relativistic electron vortex beams in a laser field, Phys. Rev. Lett.  \textbf{115}, 194801 (2015); Unified approach towards the dynamics of optical and electron vortex beams, Phys. Rev. Lett. \textbf{116}, 144801 (2016).

\bibitem{Hayrapetyan}
A. G. Hayrapetyan, O. Matula, A. Aiello, A. Surzhykov, and S. Fritzsche, Interaction of Relativistic Electron-Vortex Beams with Few-Cycle Laser Pulses, Phys. Rev. Lett. \textbf{112}, 134801 (2014); O. Matula, A. G. Hayrapetyan, V. G. Serbo, A. Surzhykov, and S. Fritzsche, Radiative capture of twisted electrons by bare ions, New J. Phys. \textbf{16}, 053024 (2014); K. van Kruining, A. G. Hayrapetyan, and J. B. G\"otte, Nonuniform currents and spins of relativistic electron vortices in a magnetic field, Phys. Rev. Lett. \textbf{119}, 030401 (2017).

\bibitem{RajabiBerakdar}
A. Rajabi and J. Berakdar, Relativistic electron vortex beams in a constant magnetic field, Phys. Rev. A \textbf{95}, 063812 (2017).

\bibitem{Barnett}
S. M. Barnett, Relativistic electron vortices, Phys. Rev. Lett. \textbf{118}, 114802 (2017).

\end{thebibliography}
\end{document}